\documentclass[8pt]{iopart}
\usepackage{iopams} 
\usepackage[latin1]{inputenc}
\usepackage{graphicx}
\usepackage{url}

\begin{document} 
 
\title[Spontaneous emission -- graphene waveguides -- surface plasmon polaritons]{Surface recoil force on dielectric nano--particles enhancement via graphene acoustic surface plasmons excitation: non--local effects consideration}

\author{Julieta Olivo}
\address{Consejo Nacional de Investigaciones Cient\'ificas y T\'ecnicas (CONICET).}
\address{Facultad de Ingenier\'ia-LIDTUA-CIC, Universidad Austral, Mariano Acosta 1611, Pilar 1629, Buenos Aires, Argentina}
\address{Departamento de F\'isica, Universidad de Buenos Aires and IFIBA, Ciudad Universitaria, Pabell\'on I, Buenos Aires 1428, Argentina}
\author{Hernan Ferrari}
\address{Consejo Nacional de Investigaciones Cient\'ificas y T\'ecnicas (CONICET).}
\address{Facultad de Ingenier\'ia-LIDTUA-CIC, Universidad Austral, Mariano Acosta 1611, Pilar 1629, Buenos Aires, Argentina}
\author{Mauro Cuevas }
\address{Consejo Nacional de Investigaciones Cient\'ificas y T\'ecnicas (CONICET).}
\address{Facultad de Ingenier\'ia-LIDTUA-CIC, Universidad Austral, Mariano Acosta 1611, Pilar 1629, Buenos Aires, Argentina}
\ead{mcuevas@austral.edu.ar}

\begin{abstract} 
Controlling  opto--mechanical interactions at sub--wavelength levels is of great importance in academic science and nano--particle manipulation technologies. This letter focuses on the improvement of the recoil  
force on nano--particles placed close to a graphene--dielectric--metal structure. The momentum conservation involving the non--symmetric excitation of acoustic surface plasmons (ASPs), via near field circularly polarized dipolar scattering, 
implies the occurrence of a huge momentum kick  on the nano--particle. 
Owing to the high wave--vector values entailed in the near field scattering process, it has been necessary to consider  
the non--locality of the graphene electrical conductivity 
to explore the influence of the scattering loss on this large--wave--vector region, which is neglected by the semi--classical model. Surprisingly, 
the contribution of ASPs to the  recoil 
force 
is 
negligibly modified when the non--local effects are incorporated  through the graphene conductivity. On the contrary, our results show that the contribution of the non--local scattering loss to this force becomes dominant when the particle is placed very close to the graphene sheet  and that it is mostly independent  of the dielectric thickness layer. Our work can be helpful for designing new and better performing large--plasmon momentum opto--mechanical structures using scattering highly dependent of the polarization for moving dielectric nano--particles.   
\end{abstract} 



\maketitle
\ioptwocol

The ability to control the state of movement of nano--particles by exchanging momentum with light is of great interest in chemical and biological applications. Besides manipulation using structured beams, which has allowed the development of modern approaches for trapping   into small regions, pushing or pulling forces or rotating nano--particles \cite{Chen-bessel,Li-bessel,Zhou-superfast}, the use of optical cavities enabling highly confined optical modes has emerged as a new degree of freedom in the last ten years. These   configurations, based on plane wave excitation of the  eigenmodes of the optical structure, have aroused an  increasing interest due to their capability for providing optical trapping beyond the diffraction limit, 
because they are less sensitive to accurate positioning and alignment of  laser beams and for not requiring translating any focused  beam to move particles \cite{Jin-hyperbolic,prb-fortuno}.   
An outstanding kind of such forces are the  recoil force acting on dielectric nano--particles coming from the metallic surface plasmons (SPs) non-symmetric excitation by plane wave incidence    \cite{petrov2016surface,ferrari2023_metal,2019}.

The advent of plasmonic materials at terahertz (THz) and infrared (IR) region, such as graphene, has aroused an  interest on moving some plasmonic force generation techniques, currently working in the visible region, to those frequency ranges \cite{OT1menos,OT0,OT1,OT2,OT22,FZC}. 
The main idea behind it is to avoid the highly power losses, resulting in heat dissipation by Joule effect limiting the performance for manipulating living biospecimens,  commonly appearing on metallic plasmons excitation \cite{Wang, Roxworthy, ferrari2022giant}. 
Since the recoil force 
on dipolar particles largely depend on momentum exchanging via surface plasmon asymmetric excitation, the improvement   of these force intensities can be reached provided that the particle be placed near a structure supporting SPs  with large wave--vector. In this way, the particle will be able to receive a huge kick, when it is compared with that provided by the  photon, as consequence of momentum exchanging with SPs. 
\begin{figure}
\centering
{\includegraphics[width=0.4\textwidth]
{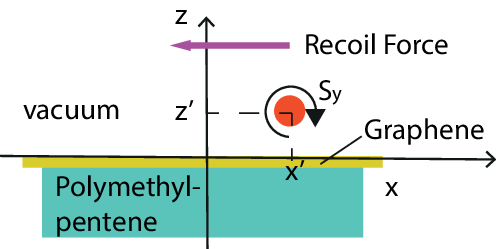}}
\vspace{-0.4cm}
\caption{Scheme of the system. A dielectric nanoparticle placed at position $\mathbf{r}'=x'\hat{x}+z'\hat{z}$, that has a rotating dipole moment $s_y$ (positive in Figure), is forced via SP scattering  on the $-x$ direction by action of a  recoil force.}
\label{sistema}
\vspace{-0.7cm}
\end{figure}
Furthermore, it is known that  graphene--dielectric--metal structures support SPs with huge wave--vectors \cite{acoustic-Liu,acoustic-Goncalves,dipole-AGP-nikitin}, 
named acoustic surface plasmons (ASPs) due to a linear dependence between frequency and wave--vector, a property that makes them ideal candidates for THz recoil force enhancement. 

The aim of this work goes in that direction. 
In virtue of the 
large wave--vector values attained 
in the scattering process, a phenomenon that is more pronounced as the nano--particle is closed to the graphene sheet,   the spacial variations of the electric field  can exceed the classical limit and, consequently,  the non--locality or spatial dispersion in electric conductivity arises \cite{kopens-NL}. In consequence, we have considered the graphene as an infinitesimally thin layer characterized  by the non--local conductivity following the Lindhard--Mermin  model \cite{peres-libro} and we compared the results with those obtained by applying the local theory. 
%
Note that, unlike  in the modern theory of optical forces, where non--local effects refer to the optomechanical manifestation of illumination in the vicinity of the particle \cite{non-local}, in this work we use this term to  represent the dispersion in the graphene conductivity. 

The proposed structure here studied  composes of a dielectric layer 
 with relative permittivity $\varepsilon_2=2.13$ (corresponding to Polymethylpentene) of thickness $d$ coated with graphene (electrical conductivity $\sigma(k,\omega)$, $\omega$ and $k$ are the angular frequency and the wave--vector, respectively) and placed on top of a metallic medium (Figure 1). A dielectric nano--particle is placed above the structure, in vacuum medium ($\varepsilon_1=1$), at position $\mathbf{r}=x' \hat{x}+y' \hat{y}+ z' \hat{z}$. 
\begin{figure}
\centering
{\includegraphics[width=0.45\textwidth]
{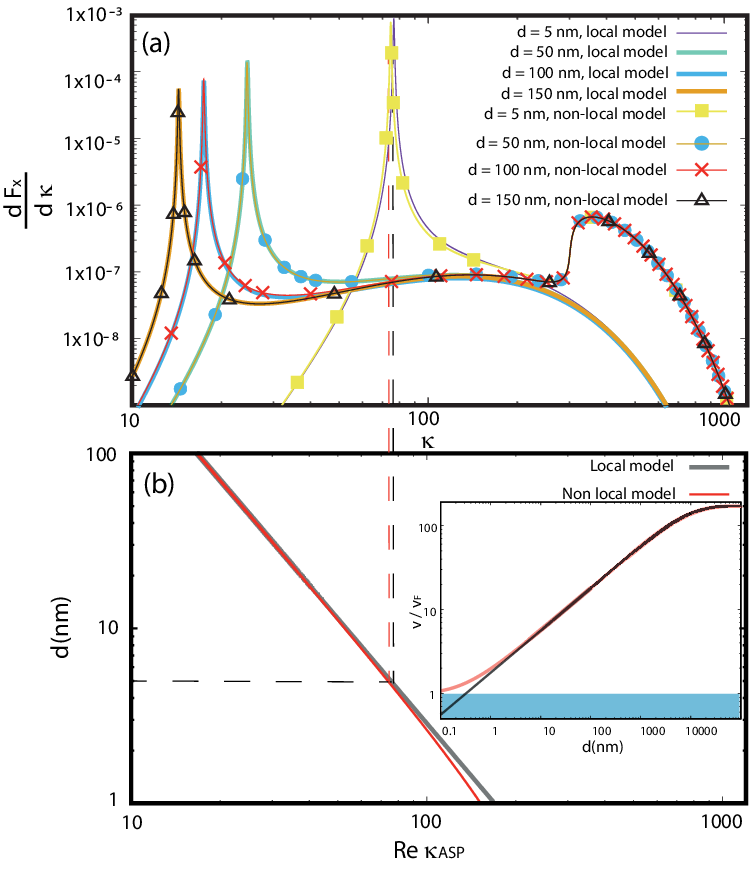}}
\caption{(a) Force Spectrum function $\frac{d} {d\kappa} F_x$ as a function of the normalized $\kappa$ wave--vector for various values of the dielectric thickness, $d=5,\,50,\,100,\,\mbox{and}\,150$ nm. The particle is placed at position $z'=150$ nm. (b) Real part of the normalized propagation constant $\Re \,\kappa_{ASP}$ as a function of $d$. In both plots (a) and (b) the calculations are carried out by using local and non--local conductivity models. The graphene  chemical potential $\mu_g=0.5$ eV and   collision frequency rate $\gamma_g=0.1$ meV.   
} 
\label{dispersion}
\end{figure}
 It is known that, under certain illumination conditions, a particle can  acquire an induced rotating dipole moment giving rise to a spin $\mathbf{s}$   induced on it. When such a particle is placed close to the surface, $z'$ near zero in our system of reference, the near scattered field      
  excites only ASPs with similar spin $\mathbf{s}$ 
  leading to ASPs traveling towards a determined direction along the graphene sheet. Therefore, a net force, which is  opposed to the ASPs propagation direction, appears as a consequence of the momentum conservation. This recoil   force is the source for the    pulling force  (a force  pulling the particle towards the source light) \cite{petrov2016surface} or lateral force (a force whose direction is perpendicular to the incident wave)  generation \cite{fortuno2019}. In both cases, the magnitude of the force can be calculated in terms of the spin $\mathbf{s}$, which  depends on induced dopole moment on the nano--particle $\mathbf{p}$, and the  geometrical and constitutive  parameters of the structure. Without loss of generality, we suppose that  the induced dipole moment is along the $+y$ direction, $\mathbf{s}=s_y \hat{y}$ with $s_y=-\frac{2\,\Im({p_x^{*}p_z})}{|p_x|^2+|p_y|^2}$ and, consequently, the optical force results in the $\pm x$ direction, where the sign $+$ or $-$ depends on the phase of the nominator, that is,
  on the phase difference between the induced dipole moments along the $x$ and $z$ directions. 
This situation could correspond to the case  
 for which the particle is illuminated by circularly polarized plane wave 
 at grazing incidence, \textit{i.e.}, with an incident wave vector almost parallel to the $y$ axis \cite{fortuno2019}, or by  linearly polarized plane wave with the  electric field contained in the incidence plane ($p$ polarization) \cite{petrov2016surface,ferrari2022giant}. 
 In such cases, the recoil force is written as \cite{fortuno2019},
\begin{eqnarray}\label{F_x}
F_x= P\, 
 \int_0^\infty d\kappa \, \frac{d\,F_x}{d\,\kappa},
\end{eqnarray}
where $\kappa=k_{||}/k_0$  is the normalized component of the propagation constant along the surface, $k_0=\omega/c$ is the photon wave--vector in vacuum, $c$
 is the vacuum speed of light,  and $P=\omega^4 |p|^2/(12\pi \varepsilon_0 c^3)$ is the power emitted by the same dipole in an unbounded medium 1 (vacuum in our case), this is,  far from any structure, and $\varepsilon_0$ is the vacuum permittivity. For a clear discussion, we 
have introduced the quantity $\frac{d\,F_x}{d\,\kappa}$ as the force spectrum function, 
\begin{eqnarray}\label{dF_x}
\frac{d\,F_x}{d\,\kappa}= \frac{3 }{4\,c}\, s_y \,  
 \Im{ \Bigg[\kappa^3 A(\kappa) \,e^{i 2   \gamma^{(1)} z'} \Bigg]},
\end{eqnarray}
where $A(\kappa) = a(\kappa) / b(\kappa)$ is the reflection coefficient for $p$ polarization with
\begin{eqnarray}\label{a_kappa}
a(\kappa) = \frac{\gamma^{(1)}}{\varepsilon_1} \left(1+ e^{-i2\gamma^{(2)} d}\right) - \frac{\gamma^{(2)}}{\varepsilon_2} \left( e^{-i2\gamma^{(2)}d} -1\right)  \nonumber \\ 
+ 
Z_0 \, \sigma \frac{\gamma^{(1)} \gamma^{(2)}}{k_0 \varepsilon_1 \varepsilon_2} \left(e^{-i2\gamma^{(2)}d}-1 \right), 
\end{eqnarray}
\begin{eqnarray}\label{b_kappa}
b(\kappa) = dfrac{\gamma^{(1)}}{\varepsilon_1} \left(1+ e^{-i2\gamma^{(2)} d}\right) + \frac{\gamma^{(2)}}{\varepsilon_2} \left( e^{-i2\gamma^{(2)}d} -1\right) \nonumber \\
+
Z_0 \, \sigma \frac{\gamma^{(1)} \gamma^{(2)}}{k_0 \varepsilon_1 \varepsilon_2} \left(e^{-i2\gamma^{(2)}d}-1 \right),
\end{eqnarray}
 $\gamma^{(j)}=\sqrt{\varepsilon_j k_0^2-k_{||}^2}$ $(j=1,2)$ and $Z_0=\sqrt{\frac{\mu_0}{\varepsilon_0}}$ is the impedance vacuum.   %
 %

In our calculations, we have chosen  $k_0=0.05\mu$m$^{-1}$ corresponding to  frequency $\omega/(2\pi) \approx 2.4$THz. Without loss of generality, we consider the
case $s_y=1$, corresponding to a circular
polarized induced dipole moment $\mathbf{p}$. 
 To analyze the main contributions to the force, in  Figure \ref{dispersion} we have plotted the force spectrum function 
 as a function of the normalized wave--vector $\kappa$ for $z'=150$nm and for different values of the dielectric thickness $d$.  As we can expect, for low values of $\kappa$ both the local and non--local calculations give similar results. In this spectral region, we observe a pronounced peak corresponding to ASPs excitation. The spectral position of these  peaks, which are the main contributions to the force (\ref{F_x}) in the low wave--vector region (low--$\kappa$ region), is largely depending on the thickness $d$, as can be seen in Figure \ref{dispersion}a. To clarify, in Figure \ref{F_x}b we have  calculated the real part of the ASP propagation constant, $\Re\,\kappa_{ASP}$, as a function of the  $d$ thickness in the same wave--vector scale as in Figure \ref{dispersion}a. To do this, we have solved the modes dispersion equation, \textit{i.e.}, the zeroes of the denominator of the  reflection coefficient $A(\kappa)$, by  using a Newton--Raphson method adapted to the complex  variable $\kappa$ \cite{Zeller2011Riemann}.  We see that the curves $\Re\,\kappa_{ ASP}(d)$ calculated using the local and non--local  conductivity 
 are almost indistinguishable, having a little difference for values of $d$ small  enough. For example, the calculated real parts of the ASP propagation constant for $d=5$nm are  $\kappa=75.4$ and $\kappa=73.7$ for the local and non--local calculation, respectively.  
 The difference between these two values, also noted by the different spectral positions of the plasmon peaks in Figure \ref{dispersion}a, agree with the fact that the spatial dispersion effects begin to manifest themselves for normalized wave--vector values of the order of 300  \cite{kopens-NL,zare-non-local}. This behaviour can be visualized in the inset in Figure \ref{dispersion}b where we have plotted the plasmon velocity $v_{ASP}=\omega/k_{ASP}=c/\kappa_{ASP}$ as a function of the thickness $d$. We observe that the velocity calculated with the local approximation begins to reduce appreciably with respect to the non--local approximation for values of $d$ lower than $\approx 5$ nm. 

On the other hand, on the high--$\kappa$ 
 region, a great difference between local and non--local calculations appears, as can be seen in Figure \ref{dispersion}a for $\kappa>300$. This condition coincides with that  as the wave--vector $\kappa$ enters on the intraband Landau damping domain, $k_{||}/k_F>\hbar\,\omega/\varepsilon_F$ where $k_{||}=k_0 \,\kappa$, $k_F$ and $\varepsilon_F$ are the Fermi wave--vector and energy, respectively.   On this spectral region, the values of the force spectral function calculated by using the non--local  conductivity exceed those   corresponding  to the local  conductivity by more than one order of magnitude. Similar results about non--local effects affecting the spectrum function on the large in--plane--wave--vectors region, but in the framework of the spontaneous emission of an atom in front of a metallic slab, have been previously reported in \cite{ford-weber}. In addition, from Figure \ref{dispersion}a  we note that, in this spectral region, there is no noticeable difference between the non--local curves for different values of $d$, which means that the spatial dispersion on the graphene conductivity 
 is mostly independent of the thickness of the dielectric layer.  

 
Next, we have calculated  the lateral force $F_x$ normalized with respect to the power $P$, \textit{i.e.}, $f=F_x/P$ in units of $N/W$. To avoid numerical 
inaccuracies coming from  singularities falling near the real $\kappa$ axis (plasmon propagation constant $\kappa_{ASP}$), by using the Cauchy theorem   we have deformed the original integration path,  which is along the real and positive $\kappa$ semi--axis, to one elliptical path passing on the negative imaginary semi--plane \cite{ferrari2022giant}. 
\begin{figure}
\centering
{\includegraphics[width=0.45\textwidth]{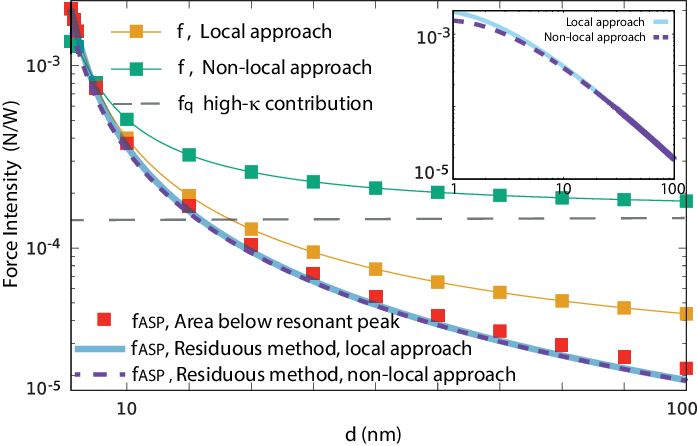}}
\caption{Normalized $f$ force as a function of the dielectric thickness $d$ calculated for both the local and non--local models. The ASP and the high--$\kappa$ contributions to the force 
 are also plotted. The nanoparticle is placed at $z'=150$ nm from the graphene sheet. The other parameters are the same as in Figure \ref{dispersion}. Inset shows the ASP contribution calculated by the local and non--local models. The horizontal and vertical axis in the inset are in log--scale.  
}
\label{fuerza_zp0p15}
\end{figure}

In Figure \ref{fuerza_zp0p15} we have plotted the normalized force together with the ASP  and the high--$\kappa$ contributions,  $f_{ASP}=F_{ASP}/P$ and $f_q=F_q/P$, as a function of the thickness $d$.  As in Figure \ref{dispersion}, the position of the particle is maintained at $z'=150$ nm.  The ASP contribution can be achieved in two ways: the first by 
calculating the area below the ASP peak in the force spectrum curves plotted in Figure \ref{dispersion}a 
into a spatial bandwidth that embraces the whole resonant peak, and the other, 
by using the residues theorem to obtain an analytical expression for the ASP contribution to the normalized force (see Supporting Information, Section S1),
\begin{eqnarray}\label{F_sp}
F_{ASP}= \frac{3 \, P}{4\,c}\, s_y \,  \pi \kappa_{ASP}^3 \frac{a(\kappa_{ASP})} {\frac{\partial b}{\partial \kappa} (\kappa_{ASP})} \,e^{- 2   k_0 \kappa_{ASP} z'},
\end{eqnarray}
where, in virtue of the fact that 
 $\kappa_{ASP} \gg 1$, we used the quasistatic approximation 
$\gamma^{(1)}\approx i k_0 \kappa_{ASP}$. 
We observe that both calculations agree reasonably  well.  
 Note that 
 the ASP contribution 
shows small differences between local and non--local calculation for values of $d$ smaller that $5$ nm (see inset in Figure \ref{fuerza_zp0p15}). This is true because the propagation constant values calculated within these approaches begin to  differentiate below $d \approx 5$ nm (see Figure \ref{dispersion}b).  
Regarding to the total force, from Figure \ref{fuerza_zp0p15} we can see that the  values calculated by using the non--local conductivity are noticeably more larger than those obtained by using the local conductivity. For instance, the non--local force exceeds  the local by about one order of magnitude for thickness $d \approx  100$ nm.  This behaviour  arises from the  fact that 
the local calculation does not takes into account the intraband Landau damping  (electron--hole excitation) which dominates the scattering loss in the high--$\kappa$ region. 

\begin{figure}
\centering
{\includegraphics[width=0.45\textwidth]
{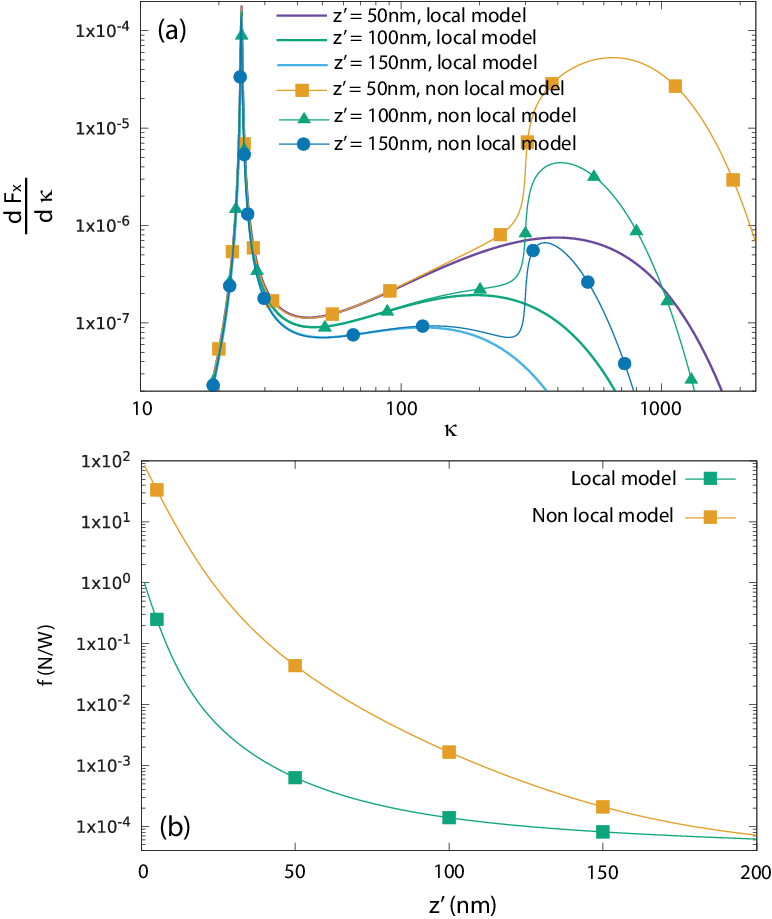}}
\caption{
(a) Force spectrum function as a functions of the normalized wave--vector for various values of $z'$, $z'=50,\,100,\,\mbox{and}\,150$ nm. (b) Normalized $f$ force as a function of the $z'$ distance. In (a) and (b) the dielectric thickness layer is fixed at $d=50$ nm.    
 The other parameters are the same as in Figure \ref{dispersion}. 
}
\label{fuerza_d=0.05}
\end{figure}

On the other hand, in the large $\kappa$ region of Figure \ref{dispersion}a the quasistatic approximation is valid. In this limit, $\kappa > 300$, 
the higher--$\kappa$ contribution to the recoil force can be approximated by 
\begin{eqnarray}\label{F_q}
F_{q}= \frac{3 \, P}{4\,c}\, s_y \frac{1}{8\, (k_0 z')^4} \, \Im\, A(\kappa>300). 
\end{eqnarray}
%
Even though the value of $\Im\,A$ depends on $\kappa$, for $\kappa>300$ we can consider that  
as a constant value, independent of $\kappa$, and 
equal to that corresponding for $\kappa=\kappa_F$.  The   calculated value of  $\Im\, A(\kappa_F) \approx 1.5 \times 10^{-3}$ (see Supporting Information, Section S2). With this value, 
the non--local contribution to the normalized recoil force $f_q  \approx 1.5 \times 10^{-4}$ (see Figure \ref{fuerza_zp0p15}). 
%

To investigate the dependence with the distance between the particle and  graphene, in Figure \ref{fuerza_d=0.05}a we have calculated the spectral function for various values of the $z'$ distance, $z'=50,\,100\, \mbox{and}\, 150$ nm. 
The dielectric thickness is fixed at $d=50$ nm. 
We see that both the plasmon and the high--$\kappa$ contributions increase as the distance $z'$ decreases, \textit{i.e.}, as the particle approaches the graphene sheet. These increments can be understood by the 
dependence of this force contributions   
 on $z'$ distance, $\mbox{exp}(-2 \, \kappa \, k_0\, z')$ and $(k_0\,z')^{-4}$ for the SP and high--$\kappa$ contributions, 
 respectively. 
 In addition,  the difference between local and  non--local  effects is noticeably increased beyond $\kappa>300$. 
In this large--$\kappa$ region, the contribution of the intraband Landau damping dominates over the plasmonic contribution and the resultant recoil  force is largely enhanced due to this effect,  as can be seen in Figure \ref{fuerza_d=0.05}b, where we plotted the normalized force using the local and non--local formulations as functions of the $z'$ distance maintaining $d=50$ nm fixed. The force values calculated with the non--local theory becomes two orders of magnitude, it is depending of $z'$, larger than those calculated by using the local theory. 
We have validated the results obtained by calculating the recoil force  on dielectric cubes using full--wave simulations (see Supporting Information, Section S3).

%
%
In conclusion,  we have compared the results referred to the optical recoil force on a dielectric nano--particle placed near a graphene--dielectric--metal planar structure using local and non--local electric conductivity approaches. We quantified the two main contributions to this force: the excitation of ASPs and intraband electron--hole pairs, this last appearing for higher values of the in--plane wave--vector. For values of the dielectric thickness layer grater than $5$ nm,  the SP contribution almost does not depend on whether the used approach is local or non--local. On the contrary, the contribution of the high wave--vector region, given rise to electron--hole pairs in the conduction band of graphene and which is taking into account in the non local approach, dominates the momentum exchange process between the nanoparticle and the electromagnetic field.


We believe that our  results can be valuable for the design of polarized--dependent opto--mechanical devices using near field scattering between  nano--particles and plasmonic interfaces, in  particular, when the particle is placed very close to the plasmonic surface where the non--local effects dominate the scattering process.

\section*{Acknowledgment}
The authors acknowledge the financial supports of Universidad Austral O04-INV00020, Agencia Nacional de Promoci\'on de la Investigac\'on, el desarrollo Tecnol\'ogico y la Innovaci\'on PICT-2020-SERIEA-02978 and Consejo Nacional de Investigaciones Científicas y Técnicas (CONICET).

\end{document}